\documentclass[showpacs,aps,prb,twocolumn]{revtex4}
\usepackage{epsfig}

\usepackage{graphicx}
\usepackage{dcolumn}
\usepackage{bm}

\begin{document}

\preprint{APS/123-QED}
  


\title{Towards a first principles description of
phonons in Ni$_{50}$Pt$_{50}$ disordered alloys: the role of relaxation}

\author{Subhradip Ghosh}
\altaffiliation{Present address: Department of Materials Science \&
                Engineering, University of Illinois at Urbana-Champaign,
                Urbana, Il 61801, USA}
\author{J. B. Neaton}
\altaffiliation{Present address: The Molecular Foundry, Materials Sciences
                Division, Lawrence Berkeley National Laboratory, 
                Berkeley, CA 94720, USA.}
\author{Armin H. Antons}
\altaffiliation{Present address: Institut of Solid State Research (IFF), 
Research Centre J{\"u}lich, 52425 J{\"u}lich, Germany.}
\author{Morrel H. Cohen}
\author{Paul L. Leath}
\affiliation{Department of Physics and Astronomy,
Rutgers, the State University of New Jersey, 
136 Frelinghuysen Road,
Piscataway,New Jersey 08854-8019, USA} 
\date{\today}
\begin{abstract}
Using a combination of density-functional perturbation theory and
the itinerant coherent potential approximation, we study 
the effects of atomic relaxation on the inelastic incoherent neutron scattering 
cross sections of disordered Ni$_{50}$Pt$_{50}$ alloys. 
We build on previous work, where 
empirical force constants were adjusted {\it ad hoc} 
to agree with experiment.  After 
first relaxing all structural parameters within the local-density 
approximation for ordered NiPt compounds,
density-functional perturbation theory is then used to compute phonon spectra, 
densities of states, and the force constants.
The resulting nearest-neighbor force constants are first compared
to those of other ordered structures of different stoichiometry, and then
used to generate the inelastic scattering cross sections within
the itinerant coherent potential approximation.
We find that structural relaxation substantially affects the computed
force constants and resulting inelastic cross sections, and that the effect
is much more pronounced in random alloys than in ordered alloys.
\end{abstract}

\maketitle

\section{\label{sec_Intr}INTRODUCTION\protect\\}

Extensive experimental studies of the lattice dynamics of random binary 
alloys over the past forty years
\cite{expt,expt1,expt2,expt3,expt4} have provided considerable insight 
into the nature of their elementary excitations. Compared 
with ordered alloys, the presence of disorder results in new 
phenomenona that depend on the impurity concentration but
also crucially on both the relative mass and size difference between 
the constituents. For large mass or size differences, 
the effect of disorder can be dramatic,
such as the appearance of sharp discontinuities (or split bands) observed 
in the dispersion. 

Among disordered alloys, the Ni$_{\rm x}$Pt$_{\rm 1-x}$ system has been
especially well-studied experimentally.\cite{dbm,ni,ni1} Structurally simple,
the system assumes variants of fcc over a wide composition
range; moreover, the components have large mass (M$_{\rm Pt}$/M$_{\rm Ni}$=3.323)
and size ratios (r$_{\rm Pt}$/r$_{\rm Ni}$=1.12). 
Further, the phonons of the endpoint 
compounds,
elemental metallic Pt and Ni, have also been well characterized both 
experimentally and theoretically.\cite{dbm,ni,ni1} The interest in these 
alloys has arisen in part because homogeneous crystals are easy to grow 
at nearly any concentration, making the system attractive for experimental 
study.

Theoretically, there have been many attempts to describe the complex nature 
of phonon excitations in random alloys. The majority were based on the 
coherent potential approximation (CPA)\cite{cpa} and its 
various generalizations.\cite{cpa1,cpa2,cpa3,cpa4,cpa5,cpa6,cpa7,cpa8}
The CPA is a single-site, mean-field theory capable of dealing only with mass disorder;
it is commonly generalized to treat both off-diagonal and 
environmental disorder.  A translationally invariant CPA based on the
augmented space formalism \cite{mook} has recently been used 
successfully to describe
the lattice dynamics of Ni$_{55}$Pd$_{45}$ and Ni$_{50}$Pt$_{50}$ random alloys.\cite{icpa} 
This formalism, known as the itinerant CPA (or ICPA),
captures the effects of both force constant {\it and} mass disorder.

A key deficiency of the application of those theories is that the 
Green's functions are
constructed from {\it empirical} force-constants. Moreover, in Ref.~\onlinecite{icpa}
empirical force constants were then themselves adjusted to fit 
frequencies and linewidths extracted from
neutron-scattering data. A better alternative to
reliance on phenomenological force constants is to compute them 
from first principles. In recent years, first-principles density-functional theory 
(DFT) has achieved the ability to predict accurately material properties
without experimental input or adjustable parameters. 
A wealth of first-principles studies has demonstrated that electronic,
vibrational, and transport properties are often extremely sensitive 
to the precise details of the atomic arrangement.\cite{rel1,rel2,rel3,rel4,rel5}
Providing a parameter-free description of the vibrational properties
of binary alloys would allow a deeper, atomic-scale understanding of the 
phenomenology.

An especially relevant atomic-scale aspect of these disordered alloys 
which can be captured by first-principles methods 
is the degree to which the atoms displace from their high-symmetry fcc sites because
of size mismatch; these atomic ``relaxations'' would
be expected to have considerable impact on the force constants.
Experimentally, non-negligible displacements are routinely found in binary 
alloys where the size difference between the constituents is large. For example, 
EXAFS measurements of CuPd alloys have shown that 1 at$\%$ Pd in Cu changes the 
Cu-Pd nearest neighbor distance to 2.560~\AA, from the 2.515~\AA~Cu-Cu 
bond distance in pure Cu.\cite{exafs1} Another EXAFS study of the 
Ni$_{\rm x}$Au$_{\rm 1-x}$ system revealed three distinct Ni-Ni, Au-Au and Ni-Au bond 
lengths.\cite{exafs2} 

Within the first-principles framework, some degree of translational 
symmetry is almost always assumed and disordered systems, such as random 
binary alloys, 
are treated approximately, usually by averaging properties over several different 
ordered atomic configurations. 
Current computational capabilities limit
the size of each configuration, or {\it supercell}, and also the number 
of configurations sampled.  However, information from one particular 
ordered compound can nonetheless provide insight into the physics 
of its disordered counterparts. 
Although the force constants of a particular ordered system are not
necessarily expected to be transferrable to a random environment, 
they should yield insight into the relative contribution of lattice composition and 
relaxation to the vibrational properties
of Ni$_{50}$Pt$_{50}$, a fundamental feature of 
these systems which has yet to be studied. 

In this article, we attempt to capture the effects of atomic relaxation 
on the inelastic incoherent neutron scattering cross section of disordered
Ni$_{50}$Pt$_{50}$ using first-principles force constants computed from 
ordered alloys as input to the ICPA. We build on previous work  
with the ICPA in which empirical force constants were adjusted in an {\it ad-hoc} 
manner to agree with experimental cross section spectra.\cite{icpa} 
With a large mass and size ratio, Ni$_{50}$Pt$_{50}$ is expected to
possess substantial atomic relaxation and large force constant ratios;
further, we have a wealth of experimental and theoretical results
for comparison.

The paper is organized as follows. In section II, we briefly describe the theoretical tools 
used in this work to connect the first-principles calculations with the random alloy 
systems. In Section III, we present phonon spectra, site-projected phonon 
densities of states, and force constant results for the ordered Ni$_{50}$Pt$_{50}$ compounds 
in pseudo-cubic and L10 structures; the force constants of 
Ni$_{25}$Pt$_{75}$ and 
Ni$_{75}$Pt$_{25}$ compounds in L12 structures are calculated for comparison. We 
then investigate the efficacy of ordered-alloy force constants in reproducing the 
spectrum of the random 50-50 alloy. Concluding remarks appear in Section IV.

\section{\label{sec_Form}Theoretical Background}

\subsection{Details of first-principles calculations}

In what follows we report computations of the ground-state properties and 
phonon spectra of Ni$_{50}$Pt$_{50}$ ordered alloys in the L10
structure. 
We consider two cases. In the first, the $c/a$ ratio of
the tetragonal L10 structure is set equal to unity and only the
volume is relaxed. We have referred to this structure above as
``pseudo-cubic'', and below it is simply referred to as ``cubic''. In the
second, the $c/a$ ratio is relaxed as well. 
Structural information and force constant data for pure Ni and Pt and ordered 
Ni$_{3}$Pt and NiPt$_{3}$ are also presented. 
We use density-functional 
theory within the local-density approximation (LDA) to relax the  
different structures.
We employ a plane-wave 
pseudopotential approach with the Perdew-Zunger parametrization \cite{pz} of 
the LDA as implemented in the PWSCF package.\cite{pwscf} 
Ultrasoft pseudopotentials\cite{dhv} are used for Ni and Pt and explicitly treat 
10 valence electrons for each species with a kinetic energy cutoff of 30 Ry. 
Nonlinear core corrections are included in the Ni pseudopotential.\cite{nlcc}
The Brillouin zone (BZ) integrations are carried out with 
Methfessel-Paxton smearing\cite{mp} using an 8$\times$8$\times$8 {\bf k}-point mesh, 
which corresponds to 70 {\bf k}-points in the irreducible wedge.
The value of the smearing parameter is 0.03 Ry. 
Hellmann-Feynman forces are calculated, and the atoms are relaxed steadily
toward their equilibrium values until the forces are less than 1 mRy/au.
These parameters are found to yield phonon frequencies converged 
to within 5 cm$^{-1}$.

Once adequate convergence is achieved for the ground-state structural properties, 
the phonon spectra are obtained from linear response using density-functional
perturbation theory (DFPT).\cite{dfpt} Within 
DFPT, the force constants are conveniently computed in reciprocal space on a finite 
${\bf q}$-point grid; to obtain the real-space force-constants, 
Fourier transforms are then performed numerically. The number of unique real-space 
force constants and their reliability depend on the density of the 
${\bf q}$-point grid: the closer the ${\bf q}$-points are spaced, the more 
accurate the force-constants of the higher neighbors will be. In this work, the 
dynamical matrix is computed on a 6$\times$6$\times$6 ${\bf q}$-point mesh 
commensurate with the ${\bf k}$-point mesh for all structures.\cite{fc}

\subsection{Itinerant Coherent Potential Approximation}

The itinerary coherent potential approximation (ICPA) is a Green's-function based 
technique for treating random substitutional alloys. It is capable of treating
off-diagonal and environmental disorder 
and therefore appropriate for the study of phonons in alloys with strong
force-constant disorder. The ICPA was developed and used to study 
Ni$_{55}$Pd$_{45}$ and 
Ni$_{50}$Pt$_{50}$ alloys.\cite{icpa} The recent results of 
Ref.~\onlinecite{icpa} demonstrate 
its advantages over the simpler CPA for understanding both the dispersion
and lifetimes of phonons in random binary alloys. However, in that work 
empirical force constants were adjusted to fit frequencies and linewidths extracted from
neutron-scattering data. 
Using DFPT to calculate the phonon frequencies and
force constants of various ordered structures allows us to make a series of
comparisons with both the empirical force-constant results and with experiment.
In doing so, we test the extent to which ordered-alloy force constants
resemble random-alloy force constants and thereby illuminate the effect of 
relaxation on force constants, as reported in the next section.

\section{Results and Discussions}

\subsection{Structural information and ground state properties}

Solid elemental nickel and platinum stabilize in the fcc structure 
under standard conditions and are mutually soluble at 
all concentrations.
When combined, they form ordered compounds at 1:3, 1:1 and 3:1 Ni:Pt atomic proportions. 
The 1:3 and 3:1 phases crystallize in the cubic L12 (Cu$_{3}$Au) structure; the 1:1 
phase takes up the L10 crystal structure \cite{structure} as shown in Fig.1. All of
these structures contain 4 atoms per unit cell.
Their electronic structures, ordering 
tendencies, and magnetic properties have been investigated in 
detail over the years.\cite{nipt}

\begin{figure}
\includegraphics[width=8.3cm]{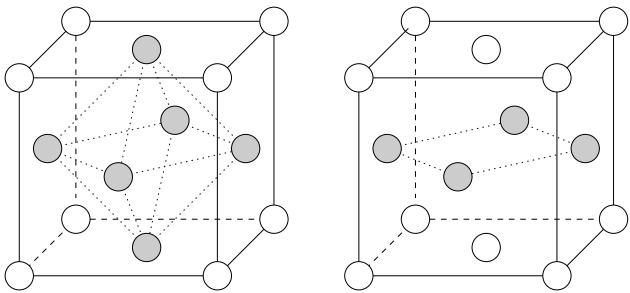}
\caption{Schematic L12 (left) and ``cubic'' (right) structures. 
For Ni$_3$Pt the white circles in the L12 structure indicate Pt atoms, 
the gray ones Ni, and vice versa for NiPt$_3$.
The white circles in the ``cubic'' structure indicate Pt atoms, 
the gray ones Ni atoms. 
In the L12 structure, atoms of each species form three dimensional next-neighbor 
networks; the ``cubic'', on the other hand, is a layered structure of Ni-Pt 
layers in the (001) direction with $c=a$.
In the L10 structure $c/a$ is free to relax, allowing an adjustment of 
the Ni-Pt layer separation.}
\end{figure}

In Table I, we report the equilibrium lattice parameters for the three compounds 
obtained from first principles calculations performed as described above
and compare them with the experimental data.\cite{exp} The LDA lattice parameters 
are 1 to 2 $\%$ less than experiment, which is within the normal range 
of error of the LDA and qualifies as very satisfactory agreement.
%
%
\begin{table}[ht]
\caption{Lattice parameters of NiPt ordered compounds calculated using DFPT
and compared to experimental values taken from Ref.~\onlinecite{exp}. 
``Cubic'' refers to the lattice constant $a$ of the NiPt structure obtained 
by fixing $c/a$ at unity.\protect\label{tab:structure1}}
\begin{tabular}{rcccc}
\hline
\hline
\vspace{-0.1in}
\hspace{0.3in} & \hspace{0.7in} & \hspace{0.7in} & \hspace{0.7in} & \hspace{0.7in}\\   
System  & $a${\rm ~(au)}  & {\rm Expt.} & {\rm $c/a$} & {\rm Expt.} \\
\hline
\hline
Ni             &   6.480   &  6.650  &       &\\
Ni$_{3}$Pt     &   6.744   &  6.892  &       &\\
L10 NiPt       &   7.136   &  7.244  & 0.934 & 0.939\\
``cubic'' NiPt &   6.99    &         &       &\\     
NiPt$_{3}$     &   7.200   &  7.320  &       &\\
Pt             &   7.394   &  7.410  &       &\\   
\hline
\hline
\end{tabular}
\end{table}

\subsection{Phonon spectra}

The phonon spectrum of the 50-50 ordered alloy in the L10  structure consists 
of twelve branches, 
three acoustic and nine optical. The results of our first-principles DFPT
calculations are shown along the high symmetry lines of the simple cubic BZ in the top 
panel of Fig.~2 for the ``cubic'' structure, i.e. L10 with $c=a$.
Results for the fully relaxed L10 structure are shown in the bottom panel of Fig.~2.
The corresponding densities of states (total and component-projected) are
shown on the right hand side of each figure.

\begin{figure*}[!]
\vbox{
\centerline{
\hbox{
\includegraphics[width=16.6cm]{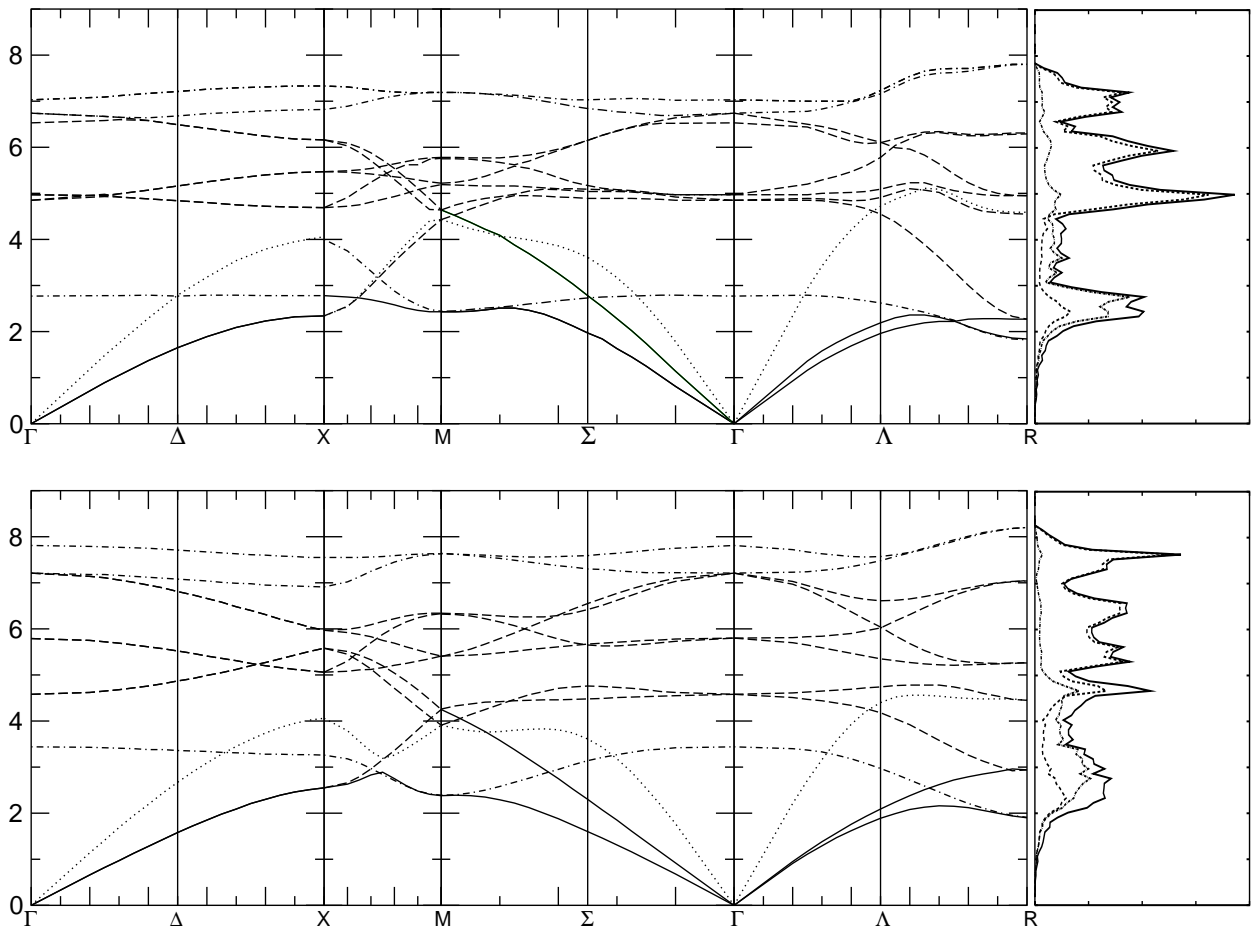}
}
}
}
\caption{Dispersion curves and phonon densities of states for
``cubic'' NiPt (top) and relaxed NiPt in the L10 structure (bottom), 
calculated from DFPT within the LDA. The leftmost panels are the dispersion
curves along high-symmetry points in the Brillouin zone. The panels on the right
are the total and component-projected densities of states.
In the panels showing the dispersion curves, the solid lines are the 
transverse acoustic modes, the dotted lines are the longitudinal acoustic modes, 
the dot-dashed lines are the longitudinal optical modes,
and the long-dashed lines are the transverse optical modes. 
In the panels showing the densities of states, the dashed lines 
are the Ni contributions,  the dotted lines the Pt contributions,
and the solid lines are the total contributions to the densities of states.}
\end{figure*}

Although many features of the ``cubic'' and L10 dispersion curves 
are similar, several key differences emerge upon close examination.
The frequencies of the optical branches in all three principal directions $\Gamma$-X,
$\Gamma$-M and $\Gamma$-R are higher in L10 than 
in the ``cubic'' arrangement. Indeed, since the unrelaxed ``cubic'' 
structure has a slightly larger volume than the L10 (by roughly 0.5\%),
it would be expected to possesses frequencies that are, in an average sense, 
smaller than those seen in the L10. Comparing the eigenvectors and the projected 
densities of states of the two structures,
we find that, for both structures, the longitudinal optical (LO) branches are heavily dominated by 
the vibrations of the lighter Ni atoms, and the acoustic modes are rich in
the motions of heavier Pt atoms, again as expected.  However, closer scrutiny of the L10
dispersion curves reveals a much more appreciable Pt contribution at intermediate frequencies
($\sim4$-$5$~THz). The enhanced Ni-Pt mixing at these frequencies is consistent with the fact that
in L10, the interplanar distance decreases from 4.9427 to 3.335 au. 

In the ``cubic'' structure, component-projected densities of states indicate low frequency 
phonons are dominated by Pt and the high frequency phonons are dominated by Ni, 
with very little evidence of available states around the middle. This stricter
separation of the Ni and Pt contributions is the signature of split-band behavior. 
Since no such split-band behavior is seen in random Ni$_{50}$Pt$_{50}$ alloy,\cite{icpa}
the force constants computed from the ``cubic'' structure would not
be expected to represent those of the random alloy.
Relaxing to L10, on the other hand, results in contributions from 
both Ni and Pt in the low-frequency acoustic branches across the Brilloun zone, 
consistent with the appearance of more states in the middle of the spectrum;
the branches are more evenly distributed in energy.
Similarly, the optical branches, also very closely spaced in the ``cubic''
structure, are seen to broaden in L10. These results illustrate how
structural relaxation modifies the vibrational spectra of the ordered alloy,
bringing it closer to that of the disordered alloy within which atomic
relaxation is less constrained.
\begin{table*}[!]
\caption{Real-space nearest neighbor force constants for the elements and
the three compounds calculated with DFPT; the empirical (emp) force constants of
Ref.~\onlinecite{icpa} for the random NiPt alloy are included for
comparison. The units are dyn cm$^{-1}$.}
\vbox{
\centerline{
\hbox{
\begin{tabular}{lcccccccr}
\hline
\hline
\vspace{-0.1in}
\hspace{0.67in} & \hspace{0.67in} & \hspace{0.67in} & \hspace{0.67in} & 
\hspace{0.67in} & \hspace{0.67in} & \hspace{0.67in} & \hspace{0.67in} &
\hspace{0.67in} $\,$\\
 & Pt & NiPt$_3$ & "cubic" NiPt &  L10 NiPt & NiPt (emp) & Ni$_3$Pt & Ni & \\
\hline
\hline
Ni-Ni  &        &        &  -4863 &   -5252 & -15587 & -11758 & -13800 & 1$xx$\\
Ni-Pt  &        & -13200 & -17418 &  -13279 & -13855 & -29762 &        & 1$xx$\\
Pt-Pt  & -27744 & -37571 & -51175 &  -40158 & -28993 &       &        & 1$xx$\\
\hline
Ni-Ni  &        &        &    318 &     498 & 436 &   253 &    149 & 1$zz$\\
Ni-Pt  &        &   2813 &   2171 &    2643 & 348 &  3044 &        & 1$zz$\\
Pt-Pt  &   5512 &   5559 &   9486 &    6623 & 7040 &       &        & 1$zz$\\
\hline
Ni-Ni  &        &        &  -5373 &  -6801  & -19100 & -13028 & -15530 & 1$xy$\\
Ni-Pt  &        & -15949 & -21885 &  -15859 & -15280 & -35310 &         & 1$xy$\\
Pt-Pt  & -30969 & -40474 & -59322 &  -45529 & -30317 &       &         & 1$xy$\\
\hline
\hline
\end{tabular}
}
}
}
\end{table*}
Table II lists the computed real-space force constants for the
elements, the ordered phases, and, for comparison, the empirical
force constants of Ref.~\onlinecite{icpa} for the disordered 50-50
alloy. We first consider the composition dependence of the force
constants of the fully relaxed structures in the sequence Pt,
NiPt$_{3}$, L10 NiPt, Ni$_{3}$Pt and Ni. All Pt-Pt and Ni-Ni
force constants increase monotonically in magnitude with Ni
concentration, the Pt-Pt and Ni-Ni nearest neighbor separations
decrease concomitantly, compressing the atoms, roughly speaking,
and causing the force-constant increase. The Pt-Pt force constants,
however, increase substantially less than the Ni-Ni force constants.
The ratio of maximum to minimum force constants ranges from 1.24 ($zz$-component)
to 1.47 ($xx$) for Pt-Pt, whereas that for Ni-Ni ranges from 2.28 ($xy$)
to 3.34 ($zz$). We note also that the composition dependence of the 
Ni-Pt force constants is weakly nonmonotonic for the $zz$ and $xy$
components with little difference in all three force constants 
between their values from NiPt$_{3}$ and L10 NiPt.

In Ref.~\onlinecite{icpa}, an excellent fit to the experimental
coherent and incoherent inelastic scattering crossections at the 50-50
composition was obtained through adjustment of the measured Ni-Ni,
Ni-Pt, and Pt-Pt nn force constants. In particular, the Ni-Ni nn force constants 
had to be reduced well below their experimental values in pure Ni
to obtain good agreement, while adjusting the Pt-Pt nn force constants was found to be
substantially less effective in modifying 
the cross sections. The Pt-Pt force constants were thus 
modified only slightly from their experimental values in pure Pt, 
as seen in Table II. (Details of the fitting strategy are
given in Ref~\onlinecite{icpa}.) This receives 
some {\it post hoc} justification given the relative
insensitivity of the Pt-Pt nn force constants to composition manifest in 
Table II and noted above. We comment further in the next section on the
implications of the sequence of force constants for the three
50-50 compositions in the table.\\

\subsection{Ordered alloy force constants as a first approximation to
the disordered alloy }

A detailed understanding of phonon excitations in disordered alloys has been impeded by
absence of detailed knowledge of the force constants. Even in the case of substitutional 
disorder, the atoms have the freedom to relax, and, due to the random 
occupation of sites, the interactions among various species are expected to
be quite different than in the ordered alloy. 

To explore the impact of lattice relaxation on the force
constants, we calculate the incoherent inelastic scattering
crossection of the random NiPt alloy using the first-principles
force constants of the unrelaxed
``cubic''structure and the fully relaxed L10 structure. The results are
shown in Fig. 3. Note that the ``cubic'' curve is considerably higher 
than the data at low frequencies, and lower at high frequencies. 
These discrepancies are substantially 
reduced by relaxation: the force constants from L10 markedly improve the agreement
between experiment and theory.
Detailed differences do remain between the L10 and the experimental curves, however.
In particular, the cross section is still higher at than the data
at low frequencies (below about 4~THz), and the high
frequency cutoff predicted from the L10 force constants is lower than measured,
a disagreement unchanged by relaxation. We attribute these
discrepancies to the constraints on relaxation enforced by the order on the nearest
neighbor separations. In the random alloy such constraints are obviously absent, and
force constants computed from larger supercells averaged over several configurations
may improve the agreement. For example, allowing the Ni-Ni nn separation to
drop below that predicted from our ordered L10 structure would increase the
Ni-Ni force constants, shifting spectral weight from lower to higher frequency in
Fig. 3 and improving
 the agreement with the empirical force constants 
in Table II. 
Similarly, the Pt-Pt nn separation can increase, which would tend to reduce the Pt-Pt 
force constants toward their empirical values in Table III. The case of the
Ni-Pt force constants is more subtle. The Pt-Pt force constants are larger in
magnitude than the Ni-Ni force constants, which suggests that the energetic cost of
compressing Pt is larger than that of expanding Ni. This suggests in turn that the
Ni-Pt separation increases in the random alloy, which would also reduce the Pt-Ni
force constants, as we find.

\begin{figure*}[!]
\vbox{
\centerline{
\hbox{
\includegraphics[width=16.6cm]{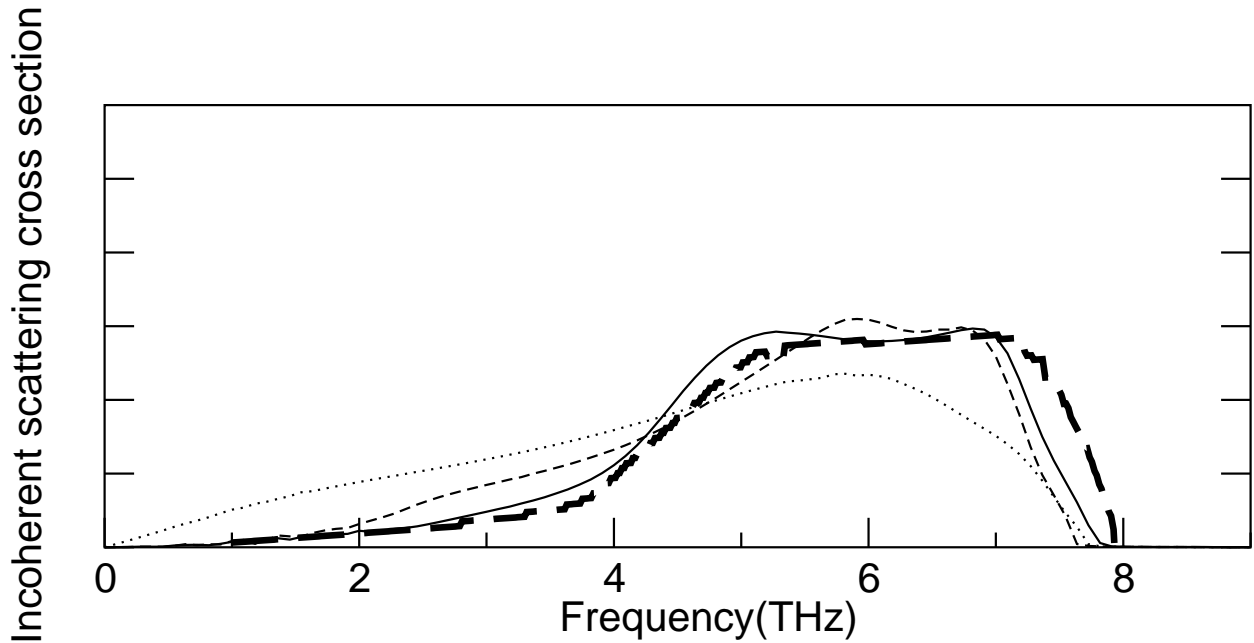}
}
}
}
\caption{Inelastic incoherent scattering cross sections for
Ni$_{50}$Pt$_{50}$ random alloy. The solid line is the experimental result,
the bold dashed line is the ICPA result with empirical force constants,
the dashed line is the ICPA result with first-principles force constants
for L10 NiPt and the dotted line is the ICPA results with first principles
force constants for ``cubic'' NiPt.}
\end{figure*}

Several other interesting trends can be seen in the force constants.
The lattice constant of the ``cubic'' NiPt alloy is 6.99 a.u.  
Due to the contraction of the ``cubic'' lattice relative to the tetragonal L10 structure,
the larger Pt atoms have less space to vibrate, resulting in a stiffening of 
the Pt-Pt interaction, as can be seen in Table II. 
In contrast, since Ni is a relatively much smaller ion
than Pt, the Ni-Ni interaction does not significantly change. For this reason we
predict too much weight in the low-frequency region
of the scattering cross section in Fig. 3. The low-frequency region is dominated 
by Pt atoms and, due to the considerable hardening of Pt-Pt compared to Ni-Ni,
the Pt-Pt interaction dominates throughout the entire frequency spectrum for 
``cubic'' NiPt. After relaxation to the L10 structure, the situation improves. The atomic
separation in the Ni and Pt planes grows (since the in-plane lattice parameter
increases), softening the Pt-Pt interactions and producing better agreement with
the experiment. These results underline the importance of capturing
relaxation effects in random environments. 

While we have captured {\it some} of the relaxation occuring in the random alloy, a deficiency
of our theory is certainly that we fail to capture {\it all} of it. For example,
we neglect randomness in the relaxation , and this may be responsible
for our underestimate of the position of the high-frequency band-edge in Fig. 3.
As mentioned, the Ni-Ni interaction is still too weak to push the upper band-edge towards higher
frequency.  It also has to be noted that the shapes and the features of the two 
scattering cross section curves obtained by ICPA using the
``cubic'' and L10 force constants have very little similarity, implying that
structural relaxation is even more pronounced in
random alloys than in ordered systems as stated above. 
The incoherent scattering cross
sections are nothing but the component-projected densities of states 
weighted by their incoherent scattering lengths and therefore carry the
signature of the component-projected densities of states. Fig. 2
suggests that though quite dissimilar in their various features, there is
more overall similarity in the densities of states curves for the ordered
structures than to that of their random counterpart.

Since the use of L10 ordered-alloy force constants improves agreement with
experiment, it is tempting to follow the same procedure 
for Ni$_{\rm x}$Pt$_{\rm 1-x}$ at $x$=0.25 and $x$=0.70, other
concentrations at which experiments have been done.
However, in the simplest ordered counterpart of
Ni$_{25}$Pt$_{75}$, a Ni atom does not have any Ni nearest-neighbors
and the Ni-Ni force constant has to be estimated empirically. Moreover,
Ni$_{70}$Pt$_{30}$ does not even have a simple ordered counterpart. 
Thus the agreement for the 50-50 case may be of limited general use.
The force constant data in Table II, however, suggest that the
Ni-Ni and Pt-Pt force constants vary approximately linearly with composition, 
following a Vegard's law of sorts for force constants. Assuming this holds 
throughout the concentration range, one could then easily interpolate to an arbitrary 
concentration and use the results as a reasonable first guess 
for force constants at that
particular concentration. On the other hand, the Ni-Pt force constants are
approximately the same for NiPt$_{3}$ and NiPt(L10), justifying a linear
interpolation for starting values for Ni$_{30}$Pt$_{70}$ alloy.

As alluded to above, another route would be to consider more complicated ordered 
structures. Zunger and collaborators have
demonstrated that a clever choice of supercell, plus 
additional relaxation can reliably reproduce optical and thermodynamic 
properties of certain binary alloys.\cite{rel5,zunger} Their supercell was constructed by arranging 
a minimal set of atoms so that the first few peaks of the radial correlation 
function matches that of the disordered alloy for a given concentration. 
Despite a plethora of works successfully using these and simpler atomic arrangements
to model the properties of alloys, to our knowledge there has yet to be a 
study assessing the ability of {\it any} of these structures to
reproduce the details of the phonon spectra of random alloys. 
In a subsequent communication, we shall examine the utility of using
such an artifical supercell to obtain force constants in a random environment.

\section{Conclusions}

In a first attempt to reach the greater goal of obtaining reliable first-
principles force constants for Ni$_{50}$Pt$_{50}$ alloys, we have
studied the vibrational properties of ordered stoichiometric compounds
of Ni and Pt using first principles DFPT. The analysis of the phonon
spectra based on the interactions between different constituents of the
alloy and their variations under different environments provides
useful insight about those interactions in random environments.
The calculated force constants of compounds of different stoichiometry
elucidate two other important features of the force constants: 
first, they provide {\it a posteriori} validation of our choice of phenomenological 
force constants \cite{icpa} for Ni$_{50}$Pt$_{50}$; 
and second, they clarify the composition-dependent variation of the force constants 
of ordered, stochiometric compounds. 
This paves the way to reasonable estimates of off-stoichiometric force constants
at arbitrary concentration. 
Most importantly, this study strongly suggests that a successful theoretical
description of the vibrational properties of random alloys must include 
the effects of atomic relaxation.

\end{document}